\documentclass[11pt,twoside]{article} 
\usepackage{asp2004}
\usepackage{epsf}
\usepackage{psfig}
\usepackage{lscape}

\begin{document} 

\title{Iron Abundance in Hydrogen-Rich Central Stars of Planetary Nebulae}
 \author{A.\ I.\ D.\ Hoffmann,$^1$ I.\ Traulsen,$^1$ T.\
 Rauch,$^{1,2}$  K.\ Werner,$^1$
 S.\ Dreizler,$^3$ and J. W.\ Kruk$^4$}
 \affil{$^1$Institut f\"ur Astronomie und Astrophysik, 
Universit\"at T\"ubingen, Sand~1, 72076 T\"ubingen, Germany \\}
 \affil{$^2$Dr.-Remeis-Sternwarte, Universit\"at Erlangen-N\"urnberg,
 Sternwartstr.7, 96049 Bamberg, Germany\\}
\affil{$^3$ Institut f\"ur Astrophysik, Universit\"at$\,$G\"ottingen,
Friedrich-Hund-Platz~1, D-37077 G\"ottingen, Germany \\}
 \affil{$^4$Department of Physics and Astronomy, The Johns Hopkins University,
 Baltimore, MD 21218, USA\\}

\begin{abstract} 
We report on an on-going analysis of high-resolution UV spectra of hot
hydrogen-rich central stars of planetary nebulae (CSPN), obtained with the
Hubble Space Telescope and FUSE. Since UV spectra of many CSPN are dominated by
Fe and Ni lines, we intend to use them as temperature indicators to check the
CSPN temperature scale we have derived earlier from CNO ionization balances.
Furthermore, the observed line strengths of heavy metals show large variations
between different objects suggesting a possible spread in abundances. We will
determine abundances of iron group elements by quantitative spectral analyses
with non-LTE model atmospheres.
\end{abstract}

\begin{figure}[t]
\plotfiddle{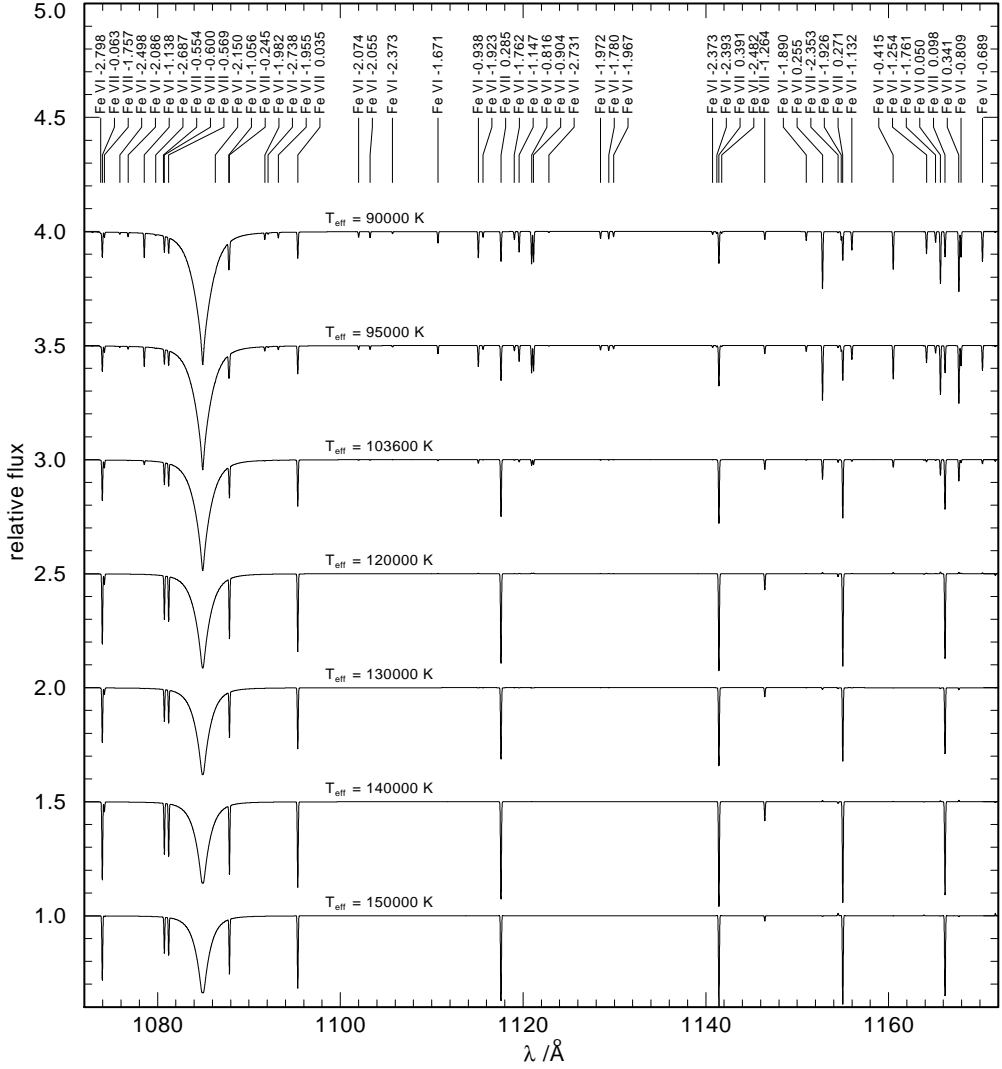}{13cm}{0}{108.0}{106.0}{-231}{-78}
\vspace{.4cm}
\caption{\hspace{-.6cm} A sequence of model spectra with increasing
  effective temperature.}
\end{figure}

\section{Introduction}
Unlike the optical spectra of hot CSPN, which are characterized by
lines of hydrogen, helium and some light metals, the ultraviolet
spectra are dominated by Fe and Ni lines (Sch\"onberner \& Drilling
1985). Their observed strengths show large variations between
different objects suggesting a possible spread in abundances. Iron
group lines are ideal temperature indicators (Fig.\,1), which is
important to set up a reliable temperature scale for the hottest
CSPN. Effective temperatures of the hottest central stars are known
with low accuracy only. As a temperature indicater one usually takes
the relative strength of neutral and ionized helium lines in optical
spectra, however, at very high temperatures neutral helium lines
disappear.

The sample of stars in our study includes seven very hot hydrogen-rich
CSPN. These are the same objects that are investigated by Traulsen et
al.\ (these proceedings) to derive temperature and gravity by
utilizing UV lines from light metals, namely C, N, and O. This sample
covers the hottest phase of post-AGB evolution
($T_{\rm eff}$\,$>$\,70\,000\,K) and
includes four objects, which have been observed with FUSE {\it and}
HST/STIS.

The FUSE spectra cover the range 910\,--\,1180\,\AA\ with a resolution
of about 0.1\,\AA. Fig.\,2 displays a section of the available
spectra.  They are ordered by increasing effective temperature
(starting from NGC\,1360 with 97\,000\,K, up to NGC\,6853 with
126\,000\,K) which becomes obvious in the shift of the iron ionisation
balance. Only lines of Fe\,\textsc{vi} and Fe\,\textsc{vii} are
labeled in this figure, and the numbers next to the identification
bars are the respective log\,\textit{gf}\hspace{1.mm} values. We also
detect lines of C, N, O and of other elements of the iron group. As
yet unidentified spectral lines are possibly absorptions of
Co\,\textsc{vi} (1139.4\,\AA), Ni\,\textsc{vi} (1096.6\,\AA,
1121.9\,\AA, 1125.4\,\AA, 1141.9\,\AA, 1145.0\,\AA, 1148.9\,\AA) and
Mn\,\textsc{vi} (1088.7\,\AA, 1128.5\,\AA).

\begin{landscape}
\begin{figure}
\plotfiddle{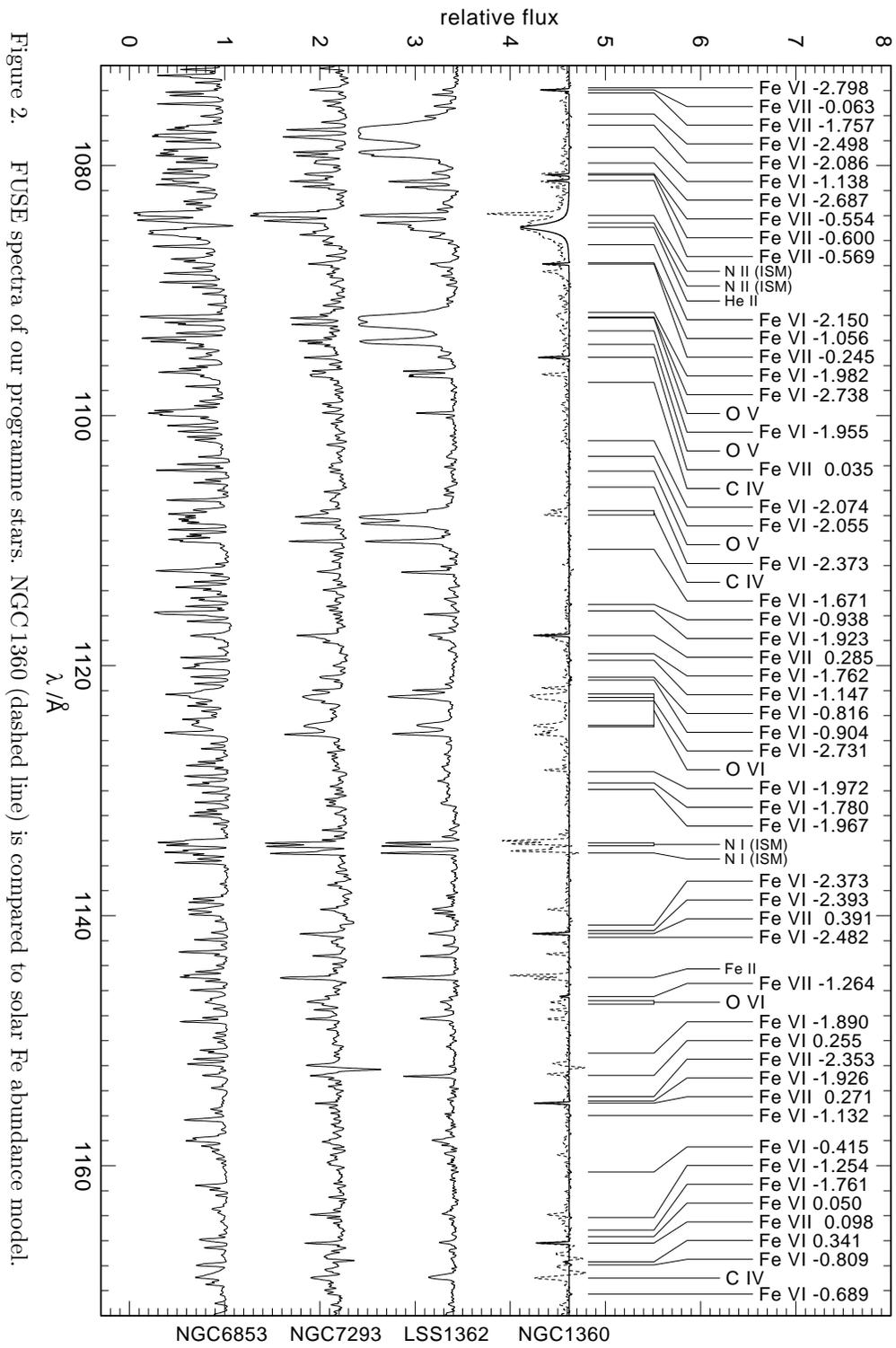}{11.cm}{270}{97.5}{97.5}{-330}{506}
\vspace{0.75cm}
\caption{FUSE spectra of our programme stars. NGC\,1360 (dashed
  line) is compared to solar Fe abundance model.}
\end{figure}
\end{landscape}
In addition to the FUSE observations, Fig.\,2 displays a model
spectrum which is plotted over the spectrum of NGC\,1360. The model
has solar abundances, and includes, besides H and He, lines from
Fe\,\textsc{vi} and Fe\,\textsc{vii}.  Temperature and gravity of the
model ($T_{\rm eff}$ $=$ 95\,000 K and log\,\textit{g} $=$ 5.50) are
close to those derived from the CNO analysis mentioned above.

Note the H$_2$ contamination of the FUSE spectra other than NGC\,1360. The very
broad troughs in LSS\,1362 are H$_2$ lines, and one can see the matching
(relatively weak) absorption in NGC\,7293.  A large fraction of the absorption
features in NGC\,6853 are from warm H$_2$, which will have to be deblended to
obtain the photospheric spectrum.

The spectra are analyzed using NLTE metal line blanketed model
atmospheres in order to determine $T_{\rm eff}$, surface gravity, and
chemical composition.  For model calculations we use the T\"ubingen
NLTE Model Atmosphere Package TMAP (Werner \& Dreizler 1999) and the
atomic data files of the iron group ions were prepared with the Iron
Opacity Interface ``IrOnIc'' (Rauch \& Deetjen 2003).  The large
number of iron lines calls for a statistical treatment of
opacities. We include data from Kurucz's (1991) line list.  The final
synthetic spectra contain only lines whose wavelength position is
accurately known from laboratory measurements (so-called POS tables of
Kurucz).  So far, all models have solar abundances and include H and
He, plus lines from Fe\,\textsc{vi}, and Fe\,\textsc{vii}.

\section{First Results}
The possibility of using the Fe\,\textsc{vi}\,/\,Fe\,\textsc{vii} ionisation
equilibrium as a temperature indicator can be seen by the disappearance of
Fe\,\textsc{vi} lines and the increasing strength of  Fe\,\textsc{vii} lines in
the spectra of models with increasing effective temperature (Fig.\,1). The
decrease of the Fe\,\textsc{vii} line strengths in the hottest model is
explained by a shift of the ionization balance from Fe\,\textsc{vii} to
Fe\,\textsc{viii}. (All models have  log\,\textit{g}\hspace{1.5mm} = 7.) Our
first calculations seem to confirm that the temperature of NGC\,1360 is indeed
lower than previously thought. The study of CNO lines arrived at a similar
result. The iron abundance in NGC\,1360 is apparently close to solar.

In future we will perform detailed model fits to the FUSE spectra and will
expand the analysis to the HST/STIS spectra as well.

\acknowledgements{T.R. is supported by DLR (grant 50 OR 0201), and J.W.K by the 
FUSE project, which is funded by NASA contract NAS5-32985.}

\end{document}